\documentclass[reprint, showpacs,
preprintnumbers,
amsmath,amssymb,
aps, floatfix,
superscriptaddress,
prd, nofootinbib]{revtex4-1}

\usepackage{graphicx}% Include figure files
\usepackage{dcolumn}% Align table columns on decimal point
\usepackage{bm}% bold math
\usepackage{hyperref}% add hypertext capabilities
\usepackage[mathlines]{lineno}% Enable numbering of text and display math
\usepackage{xcolor}
\usepackage{import}
\usepackage{lipsum}
\begin{document}
\title{Constraints On Covariant WIMP-Nucleon Effective Field Theory Interactions from the First Science Run of the LUX-ZEPLIN Experiment}

% 1 
\author{J.~Aalbers}
\affiliation{SLAC National Accelerator Laboratory, Menlo Park, CA 94025-7015, USA}
\affiliation{Kavli Institute for Particle Astrophysics and Cosmology, Stanford University, Stanford, CA  94305-4085 USA}

% 2 
\author{D.S.~Akerib}
\affiliation{SLAC National Accelerator Laboratory, Menlo Park, CA 94025-7015, USA}
\affiliation{Kavli Institute for Particle Astrophysics and Cosmology, Stanford University, Stanford, CA  94305-4085 USA}

% 3 
\author{A.K.~Al Musalhi}
\affiliation{University College London (UCL), Department of Physics and Astronomy, London WC1E 6BT, UK}

% 4 
\author{F.~Alder}
\affiliation{University College London (UCL), Department of Physics and Astronomy, London WC1E 6BT, UK}

% 5 
\author{C.S.~Amarasinghe}
% 6 
\affiliation{University of California, Santa Barbara, Department of Physics, Santa Barbara, CA 93106-9530, USA}
\affiliation{University of Michigan, Randall Laboratory of Physics, Ann Arbor, MI 48109-1040, USA}

% 7 
\author{A.~Ames}
\affiliation{SLAC National Accelerator Laboratory, Menlo Park, CA 94025-7015, USA}
\affiliation{Kavli Institute for Particle Astrophysics and Cosmology, Stanford University, Stanford, CA  94305-4085 USA}

% 8 
\author{T.J.~Anderson}
\affiliation{SLAC National Accelerator Laboratory, Menlo Park, CA 94025-7015, USA}
\affiliation{Kavli Institute for Particle Astrophysics and Cosmology, Stanford University, Stanford, CA  94305-4085 USA}

% 9 
\author{N.~Angelides}
\affiliation{Imperial College London, Physics Department, Blackett Laboratory, London SW7 2AZ, UK}

% 10 
\author{H.M.~Ara\'{u}jo}
\affiliation{Imperial College London, Physics Department, Blackett Laboratory, London SW7 2AZ, UK}

% 11 
\author{J.E.~Armstrong}
\affiliation{University of Maryland, Department of Physics, College Park, MD 20742-4111, USA}

% 12 
% ...

\collaboration{LZ Collaboration}
\date{\today}

\begin{abstract}
The first science run of the LUX-ZEPLIN (LZ) experiment, a dual-phase xenon time project chamber operating in the Sanford Underground Research Facility in South Dakota, USA, has reported leading limits on spin-independent WIMP-nucleon interactions and interactions described from a non-relativistic effective field theory (NREFT).
Using the same 5.5~t fiducial mass and 60 live days of exposure we report on the results of a relativistic extension to the NREFT.
We present constraints on couplings from covariant interactions arising from the coupling of vector, axial currents, and electric dipole moments of the nucleon to the magnetic and electric dipole moments of the WIMP which cannot be described by recasting previous results described by an NREFT.
Using a profile-likelihood ratio analysis, in an energy region between 0~keV$_\text{nr}$ to 270~keV$_\text{nr}$, we report 90\% confidence level exclusion limits on the coupling strength of five interactions in both the isoscalar and isovector bases.
\end{abstract}

\maketitle

\section{\label{sec:introduction}Introduction}
The current generation of dark matter (DM) direct detection experiments, searching for weakly interacting massive particles (WIMPs), such as LZ~\cite{LZ:Experiment_2020}, XENONnT~\cite{XenonNT:WS_2023} and PandaX~\cite{PandaX4T:SI2023}, have already probed a large parameter space for WIMPs.
These experiments have typically focused on spin-independent (SI) and spin-dependent (SD) WIMP-nucleon interactions with WIMP masses of a few GeV/$c^{2}$ to tens of TeV/$c^{2}$. 
The recent null results from both LZ~\cite{LZ:SR1WS_2022} and XENONnT~\cite{XenonNT:WS_2023} motivate the need to investigate other models.

Using an effective field theory (EFT), such as that developed by Fan \textit{et al.}~\cite{Fan_2010} and Fitzpatrick \textit{et al.}~\cite{Fitzpatrick:EFT}, it is possible to probe a wide variety of dark matter interactions and parameters in a model-independent way. 
It is possible to produce a complete set of effective operators that describe the possible interactions between WIMPs and standard model (SM) particles, making it an attractive way to increase the potential sensitivity of direct detection experiments beyond the standard SI and SD interactions.
A non-relativistic effective field theory (NREFT) framework has already been used to probe some of the potential interactions, such as in LUX~\cite{LUX:EFTR4_2021}, XENON1T~\cite{Xenon1t:2vec_2019}, PandaX-II~\cite{PandaX2:SD_EFT_2019}, and LZ~\cite{LZ:SR1_NREFT_2023}.
The operators in this NREFT can be mapped onto covariant Lagrangians, via a non-relativistic reduction of the relativistic fields, allowing for more complex interactions to be studied~\cite{Fan_2010, Fitzpatrick:EFT}.

In this letter, we perform a search for signals arising from five covariant SD Lagrangians that describe possible interactions between WIMPs and nucleons, which can only be mapped onto NREFT operators via non-relativistic reduction.
We analyze data taken by the LZ experiment during its first science run using an extended energy window, previously described in Ref.~\cite{LZ:SR1_NREFT_2023}, and perform a statistical analysis to constrain the coefficients associated with each Lagrangian.

\section{\label{sec:theory}Theory}
\begin{table*}
    \centering
    \begin{ruledtabular}
        \begin{tabular}{cccc}       
            j & $\mathcal{L}_\mathrm{int}^j$ & $\Sigma_i c_i\mathcal{O}_i$ & WIMP-Nucleon Interaction \\ \hline
                    6 & $\bar{\chi} \gamma^{\mu} \chi \bar{N} i \sigma_{\mu \alpha} \frac{q^\alpha}{m_M} N $ & $\frac{\vec{q}^2}{2 m_N m_M} \mathcal{O}_1-2 \frac{m_N}{m_{\mathrm{M}}} \mathcal{O}_3 +  2 \frac{m_N^2}{m_M{m_\chi}}\left(\frac{q^2}{m_N^2} \mathcal{O}_4-\mathcal{O}_6\right) $ &Ve-MM\\
            9 & $\bar{\chi} i \sigma^{\mu \nu} \frac{q_\nu}{m_M} \chi \bar{N} \gamma_\mu N$ & $\frac{\vec{q}^2}{2m_\chi m_M}\mathcal{O}_1 + \frac{2m_N}{m_M}\mathcal{O}_5 - 2\frac{m_N}{m_M}(\frac{\vec{q}^2}{m^2_N}\mathcal{O}_4 - \mathcal{O}_6)$ & MM-Ve \\
            10 & $\bar{\chi} i \sigma^{\mu \nu} \frac{q_\nu}{m_M} \chi \bar{N} i \sigma_{\mu \alpha} \frac{q^\alpha}{m_M} N$ & $4(\frac{\vec{q}^2}{m^2_M}\mathcal{O}_4 - \frac{m^2_N}{m^2_M}\mathcal{O}_6)$ & MM-MM \\
            12 & $i\bar{\chi} i \sigma^{\mu \nu} \frac{q_\nu}{m_M} \chi \bar{N} i \sigma_{\mu \alpha} \frac{q^\alpha}{m_M} \gamma^{5} N$  &$-\frac{m_N}{m_\chi} \frac{\vec{q}^2}{m_{\mathrm{M}}^2} \mathcal{O}_{10}-4 \frac{\vec{q}^2}{m_{\mathrm{M}}^2} \mathcal{O}_{12}-4 \frac{m_N^2}{m_{\mathrm{M}}^2} \mathcal{O}_{15}$ & MM-ED \\
            18 & $i \bar{\chi} i \sigma^{\mu \nu} \frac{q_\nu}{m_M} \gamma^5 \chi N i \sigma_{\mu \alpha} \frac{q^\alpha}{m_M} N $ & $\frac{\vec{q}^2}{m_{\mathrm{M}}^2} \mathcal{O}_{11}+4 \frac{m_N^2}{m_{\mathrm{M}}^2} \mathcal{O}_{15}$  & ED-MM \\
        \end{tabular}        
    \end{ruledtabular}
    \caption{Interactions considered in this analysis, with $j$ referring to the numerical index of the specific Lagrangian, out of the total possible 20. 
    For each Lagrangian, the relation to $\mathcal{O}$ is given in addition to the interaction that is generated.
    Lagrangians: vector (Ve), electric dipole (ED) and magnetic moment (MM)}
    \label{tab:nr_eft_lagrangians2}
\end{table*}
SD interactions between the WIMP and nucleon can be comprised of WIMP magnetic and electric dipole moments, axial-vector interference terms, and tensor interactions in addition to the standard SD physics previously studied. 
We define \textit{dimension} as 4~+ number of powers of $m_{M}$ in the denominator of the covariant interactions seen in~\autoref{tab:nr_eft_lagrangians2}, where $m_{M}$ is a normalization parameter introduced to normalize interaction to a dimensionless value. 
In this analysis, we focus on spin-1/2 WIMPs where interactions with the xenon nucleus are of dimension 5 or higher, following what has been described by Anand \textit{et al.}~\cite{Anand:MathematicaEFT}. 
At this energy scale, WIMP-nucleon interactions are constructed from the available bilinear products of scalar and four-vector interactions resulting in $2^2+4^2 = 20$ separate interaction Lagrangians.
These consist of 6 main interaction types: Scalar, Pseudoscalar, Vector, Axial Vector, Magnetic Moment, and Electric Dipole Moment interactions. 
In this analysis, we only consider Lagrangians that incorporate Electric Dipole (ED), Covariant Vector (Ve), or a Magnetic Moment (MM) coupling component in the interaction, shown in the right-most column of \autoref{tab:nr_eft_lagrangians2}.
These interactions give insight into the potential millicharged nature of the WIMP~\cite{Agrawal:2021dbo}. 
In this case, the WIMP can be composed of multiple charged particles, or be a fundamental particle with charge itself. 

A non-relativistic reduction is done between the covariant Lagrangian and the NREFT operators by replacing the spinors in the fields with the low-momentum counterparts following the prescription in Ref.~\cite{Anand:MathematicaEFT}. 
When using the NREFT, we consider a 4-body covariant interaction between the WIMP and the nucleon described by a Lagrangian
\begin{equation}
    \mathcal{L}^j_{int} = d_j\bar{\chi}\mathcal{O}_\chi^j \chi \bar{N}\mathcal{O}_N^j N,
    \label{eq:general_LD}
\end{equation}
where $j$ is an index of the Lagrangian and $d_j$ is the dimensionless coupling to be determined by the experiment that measures the effective strength or size of the interaction. 
$\bar{\chi}$, $\chi$, $\bar{N}$, and $N$ represent the non-relativistic fields of the dark matter candidate particle and nuclear targets.
$\mathcal{O}_\chi$ and $\mathcal{O}_N$ are the non-relativistic WIMP and nucleon operators described in Ref.~\cite{Anand:MathematicaEFT}. 
To write our covariant interaction in terms of this NREFT we can take for example $\mathcal{L}_{10}$: 
\begin{equation}
    \mathcal{L}_{10} = \bar{\chi} i \sigma^{\mu \nu} \frac{q_\nu}{m_M} \chi \bar{N} i \sigma_{\mu \alpha} \frac{q^\alpha}{m_M} N.
    \label{eq:lagrangian10}
\end{equation}
The leading terms come from the spatial components, so we can make the following transformations: $\gamma^{\mu} \rightarrow \gamma^{i}$ and $\sigma^{\mu} \rightarrow \sigma^{i}$. 
The relationship between particle spin and the Pauli matrix is defined as $\sigma^{i} = 2S^{i}$, so we can transform \autoref{eq:lagrangian10} into
\begin{equation}
    \mathcal{L}_{10} = 4(\frac{\vec{q}}{m_M} \times \vec{S_{\chi}}) \cdot (\frac{\vec{q}}{m_M} \times \vec{S_{N}}).
    \label{eq:lagrangian10_reduced}
\end{equation}
From Ref.~\cite{Anand:MathematicaEFT} we can retrieve that $\mathcal{O}_{4} = \vec{S_{\chi}}\cdot \vec{S_{N}}$ and  $\mathcal{O}_{6} = (\vec{S_{\chi}} \cdot \frac{\vec{q}}{m_N})(\vec{S_{N}} \cdot \frac{\vec{q}}{m_N})$ and rearrange \autoref{eq:lagrangian10_reduced} to find the reduced form of the interaction as
\begin{equation}
     \mathcal{L}_{10} = 4(\frac{\vec{q}^2}{m^2_M}\mathcal{O}_4 - \frac{m^2_N}{m^2_M}\mathcal{O}_6).
    \label{eq:lagrangian10_operator}
\end{equation}
We adopt this non-relativistic reduction for our analysis as it enables us to use existing nuclear shell model calculations when computing recoil spectra. 
Similar steps are taken to relativistically match non-linear combinations of the NREFT operators to the other covariant interactions of interest. 
The reduced Lagrangians written in terms of the operators can be seen in the center column of \autoref{tab:nr_eft_lagrangians2}.

None of the interactions presented in this analysis can be obtained from simple linear reordering of NREFT operators.
This is because in the reduced form, each operator term contains differing dependence on momentum transfer $\vec{q}$.
Differing powers of $\vec{q}$ prevent obtaining the Lagrangian result from the limits determined by previous work finding coefficients for each operator alone, as in Ref.~\cite{LZ:SR1_NREFT_2023}. 
This work therefore probes dark matter interactions beyond that of determining operator coefficients in isolation.
 
Interactions are normalized to a dimensionless value by including the term $m_M$ in the denominator of all momentum terms. 
The $m_M$ term is set equal to the nucleon mass $m_N$, normalizing the WIMP and nucleon momentum to the nucleon scale, as this is the natural scale for a theory dealing with nucleon interactions.
This choice allows us to extract information such as the size of the WIMP magnetic or electric dipole moments from the measurement of the dimensionless coupling parameter $d_j$.

\begin{figure}[h]
    \centering
   \includegraphics[width=0.44\textwidth, trim=8pt 0pt 0pt 0pt, clip]{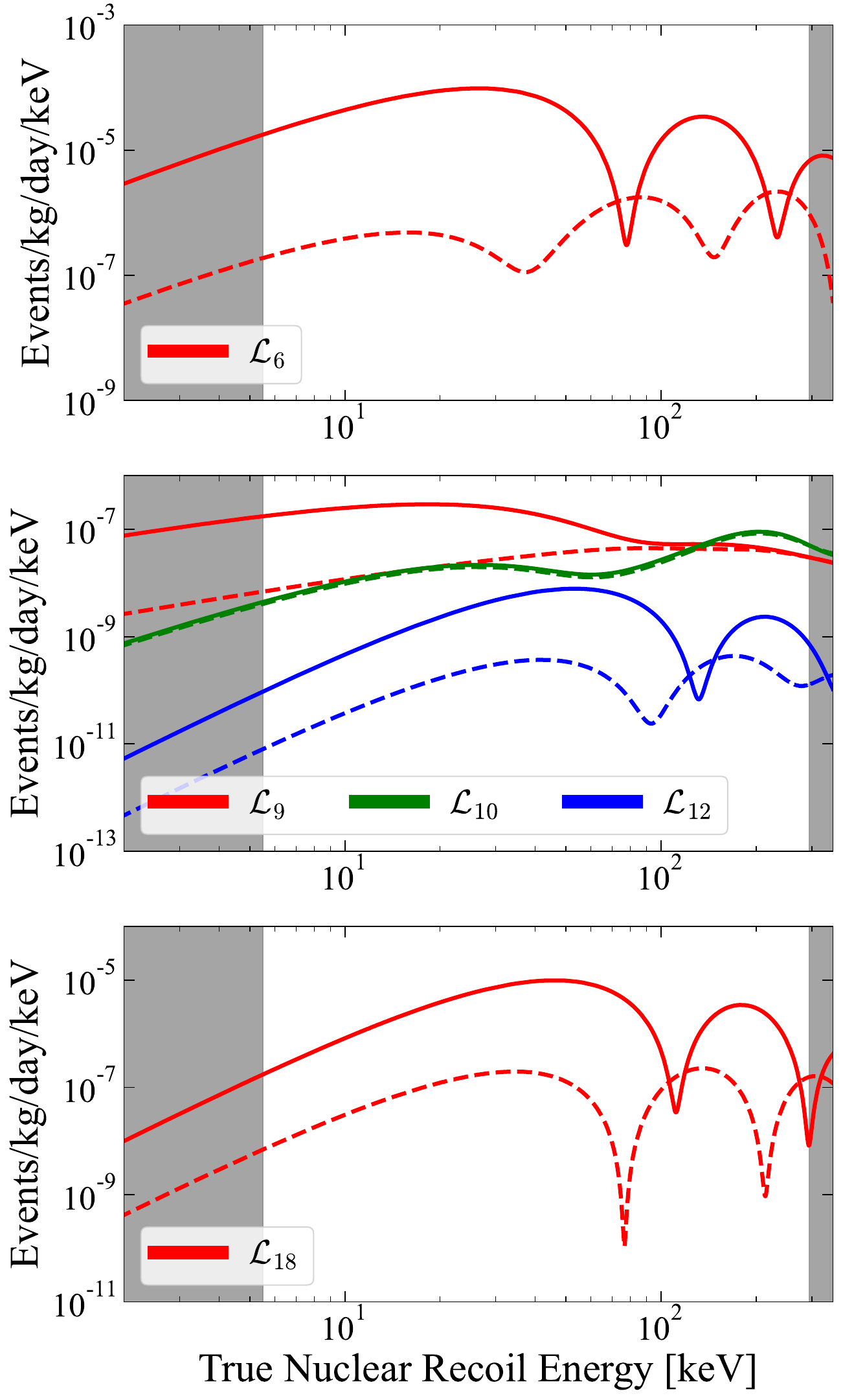}
    \caption{Differential recoil spectra from the five covariant WIMP-nucleon Lagrangians considered in this analysis. 
    Shown are the isoscalar (solid line) and isovector (dashed line) for a 1000~GeV/c$^2$ WIMP.
    The Lagrangians are categorized by the WIMP interaction: vector (top), magnetic moment (middle), and electric dipole (bottom) interactions.
    The spectra were generated with a dimensionless coupling strength of unity.
    The shaded gray regions indicate the energies at which the detection efficiency is below 50\% after all data analysis cuts have been applied (as described in Ref.~\cite{LZ:SR1_NREFT_2023}).
    }
    \label{fig:recoil_spectra}
\end{figure}

\section{\label{sec:detector}LZ Detector}
The LZ experiment, located in the Davis Campus of the Sanford Underground Research Facility, in South Dakota, USA, is centered around a low-background dual-phase time project chamber (TPC) detector containing 7~tonnes of liquid xenon (LXe) in the sensitive volume~\cite{LZ:Experiment_2020,LZ:TDR_2017}.
The cylindrical TPC is equipped with an array of photomultiplier tubes (PMTs) at the top and bottom of the detector.
These PMTs detect the energy depositions in the detector that typically make two signals. 
The first is prompt scintillation light (S1) and the second is a delayed signal (S2) caused by the electroluminescent light that occurs when electrons reach the top of the detector due to an electric field. 
The TPC is surrounded by an active LXe Skin veto detector that tags gamma-ray photons entering or exiting the TPC.
Enclosing the entire cryostat is the Outer Detector (OD), which is composed of 17~tonnes of Gd-loaded liquid scintillator and 238~tonnes of ultra-pure water to detect neutrons and muons.

The dataset used for this analysis was collected between December 2021 and May 2022 and corresponds to a total livetime of 60 days with a fiducial mass of 5.5~tonnes.
The detector condition during this period is detailed in Refs.\cite{LZ:SR1WS_2022, LZ:SR1_NREFT_2023}.
The validation of the response of the detector to nuclear recoil (NR) and electron (ER) recoil events is performed using calibration data and NEST 2.3.7~\cite{NEST:paper_2022, NEST:paper_2023} as outlined in Ref.~\cite{LZ:SR1_NREFT_2023}.
The position-corrected S1 and S2 (S1$_c$ and S2$_c$) scaling factors were $g_1=0.114\pm0.002$~phd/photon and $g_2=47.1\pm1.1$~phd/electron.

\section{\label{sec:analysis}Analysis}
The couplings of each Lagrangian are explored using the first science run of LZ in an extended energy region, as previously used in Ref.~\cite{LZ:SR1_NREFT_2023}.
An unbinned frequentist profile likelihood ratio test is performed between background and signal plus background in the S1$_c$, log$_{10}$(S2$_c$) observable space. 
S1$_c$ is constrained between 3 and 600 photoelectrons (phd) and log$_{10}$(S2$_c$)$\leq$4.5; mirroring the range used in Ref.~\cite{LZ:SR1_NREFT_2023}.

The backgrounds in the dataset, estimated in Ref.~\cite{LZ:SR1_NREFT_2023}, are dominated by a flat-ER component comprised of ${}^{212}$Pb, ${}^{214}$Pb, and ${}^{85}$Kr.
Other contributions to the ER background are ${}^{37}$Ar, ${}^{124}$Xe, ${}^{127}$Xe, ${}^{136}$Xe, ${}^{125}$I, solar neutrinos and Compton scatters from detector components.
The NR backgrounds considered are from ${}^{8}$B coherent neutrino-nucleus scattering and the scattering of neutrons originating from detector materials.
The best-fit values for each is less than one~\cite{LZ:SR1_NREFT_2023}.
Additionally, uncorrelated S1 and S2 pulses that may pair into accidental single scatter pairs are considered.

The recoil spectrum for each Lagrangian is generated using WimPyDD~\cite{Jeong_2022} with modified Xe one-body nuclear density matrices to take into account more up to date calculations~\cite{MENENDEZ2009139, haxton_unpublished}.
Following the convention set in Ref.~\cite{DM_parameters:BAXTER2021_Conventions}, the WIMP velocity distribution, $f(v)$, is described by the Standard Halo Model with $\vec{v}_\circledast$~=~(11.1, 12.2, 7.3)~km/s (solar peculiar velocity)~\cite{Schoenrich:Local_kinematics}, $\vec{v}_0$~=~(0, 238, 0)~km/s (local standard of rest velocity)~\cite{DM_parameters:galaxy_context_rest_velocity_1, DM_parameters:galaxy_context_rest_velocity_2} and $v_{\text{esc}}$~=~544~km/s (galactic escape speed)~\cite{DM_parameters:RAVE_survey_escape_velocity}. 
The local DM density, $\rho_0$, is taken as 0.3~GeV/cm$^3$~\cite{DM_parameters:LEWIN199687_DM_density}.
\hyperref[fig:recoil_spectra]{Figure~\ref*{fig:recoil_spectra}} shows the differential rate spectra for a 1000~GeV/c$^2$ WIMP-nucleon isoscalar interaction for each Lagrangian considered in this analysis.
We consider both isoscalar and isovector bases to allow future comparison with experiments with potentially different target nuclei.

\begin{figure}[h]
    \centering
   \includegraphics[width=0.44\textwidth, trim={8 5 5 5}, clip]{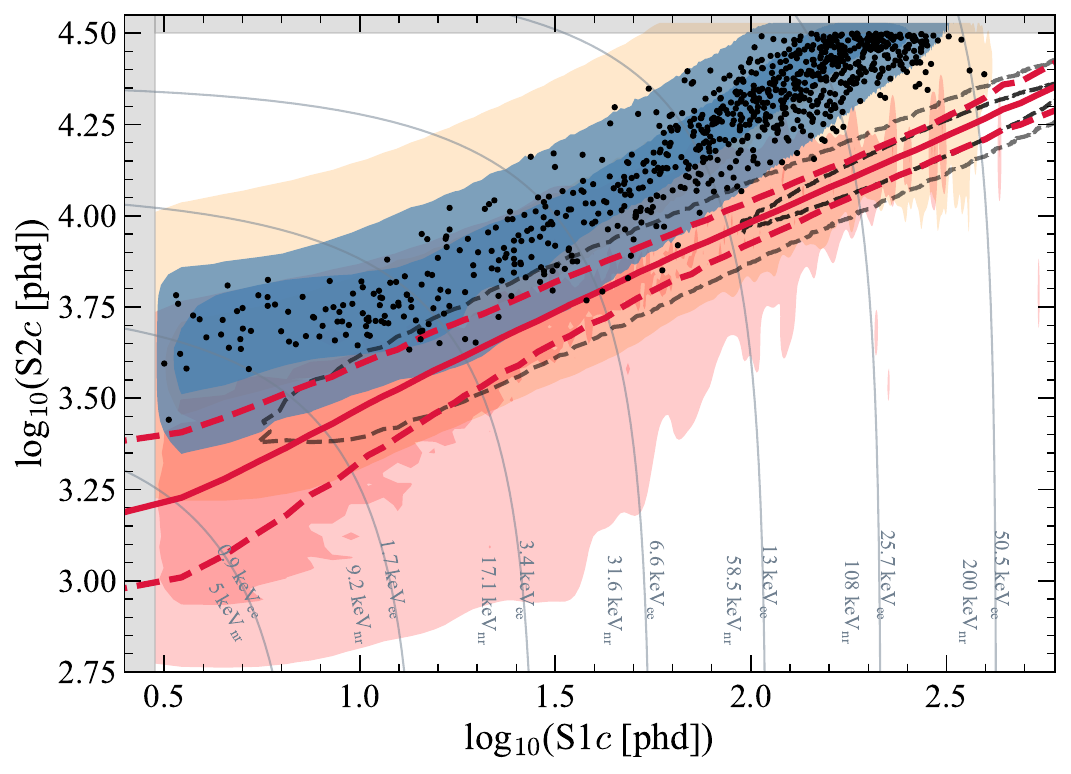}
    \caption{The final high energy WIMP-search data after all cuts in \{log$_{10}$(S1$c$), log$_{10}$(S2$c$)\} space.
    The contours that enclose 1$\sigma$ (dark) and 2.5$\sigma$ (light) regions represent the following models: 
    the shaded red region indicates neutrons originating from detector materials, the shaded orange region indicates Compton scatters from detector components, the blue region is the combined representation of all other ER models (${}^{214}$Pb, ${}^{212}$Pb, ${}^{85}$Kr, ${}^{37}$Ar,  ${}^{125}$I, ${}^{124}$Xe, ${}^{127}$Xe, ${}^{136}$Xe, and $\nu$ ER) and the black dashed lines show a 1000 GeV/c$^2$ $\mathcal{L}_6$ isoscalar signal model. 
    The solid red line corresponds to the NR median, while the red dotted lines represent the 10\%~--~90\% percentiles.
    The model contours are produced with a linear scale for S1$_c$ prior to being plotted on a log-scale and take into account all the efficiencies used in the analysis. 
    Contours of constant recoil energy have been included as thin gray lines. 
    Grayed regions at the left and top of the plot indicate parameter space outside the energy ROI.
    }
    \label{fig:models_with_data}
\end{figure}

\hyperref[fig:models_with_data]{Figure~\ref*{fig:models_with_data}} shows the \{log$_{10}$(S1$c$), log$_{10}$(S2$c$\} distribution of the 835 events which pass all selections, along with contours representing a 1000 GeV/c$^2$ $\mathcal{L}_6$ isoscalar signal model (representative of signal models that peak at non-zero energy), and the background model.

\section{\label{sec:results}Results}
\begin{figure*}
    \includegraphics[width=0.6\columnwidth]{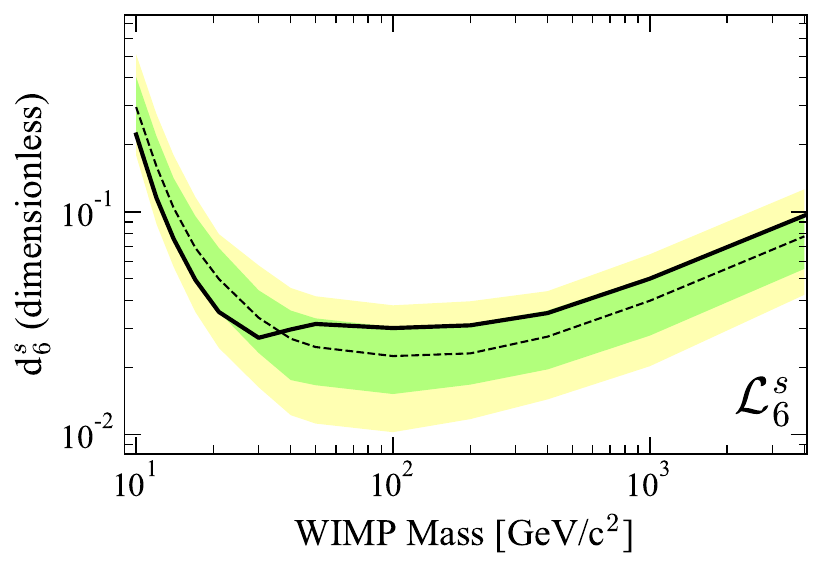}
    \includegraphics[width=0.6\columnwidth]{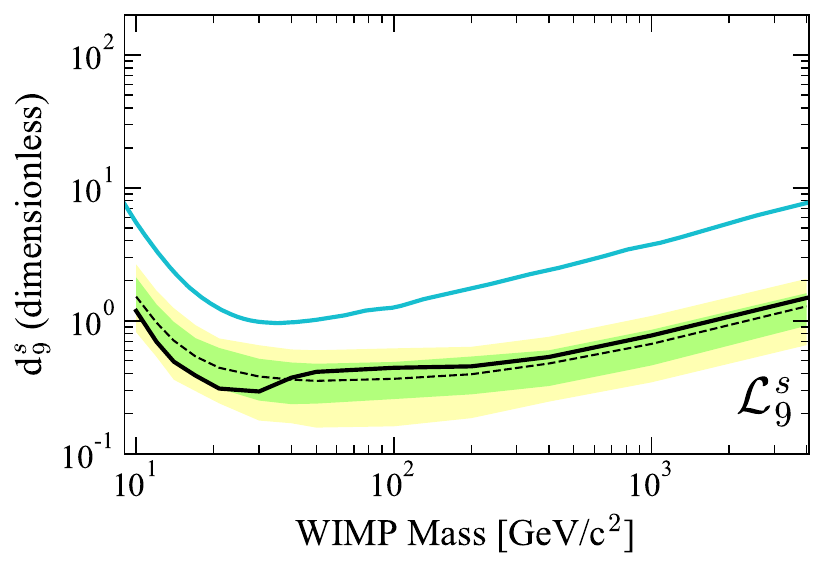}
    \includegraphics[width=0.6\columnwidth]{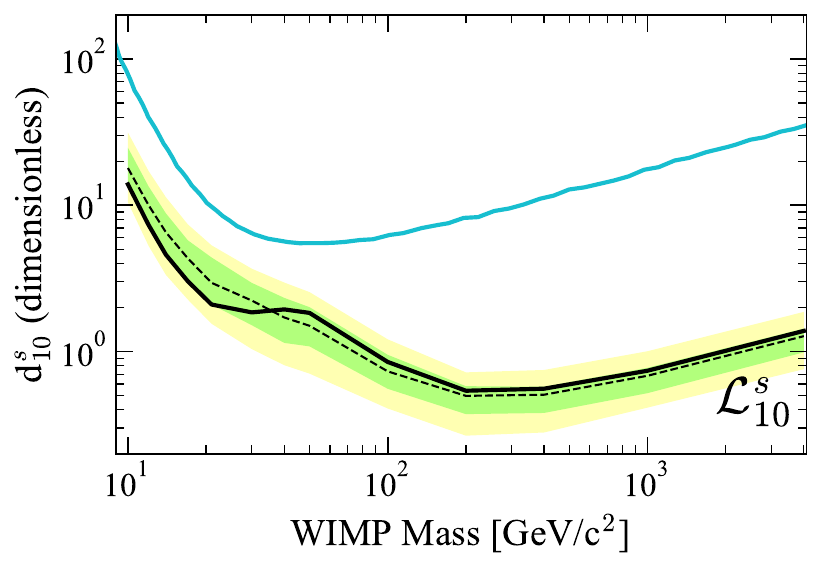}
    \includegraphics[width=0.6\columnwidth]{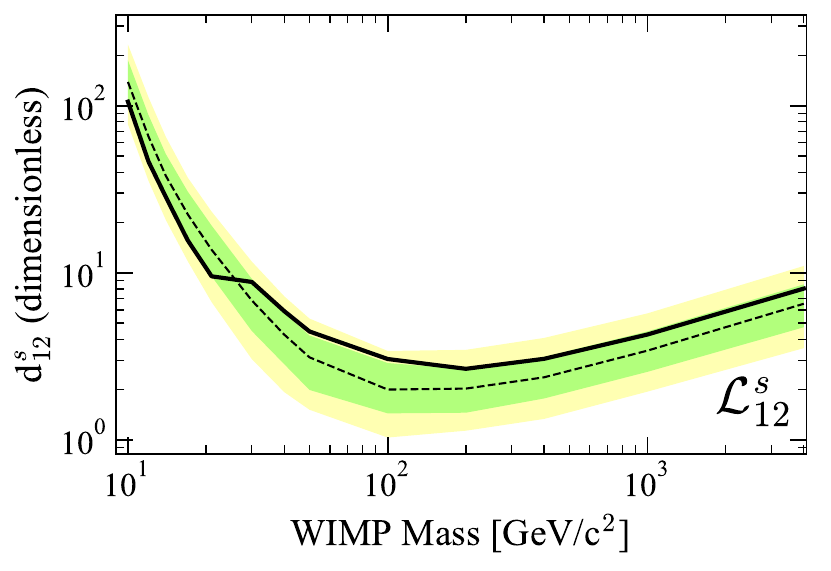}
    \includegraphics[width=0.6\columnwidth]{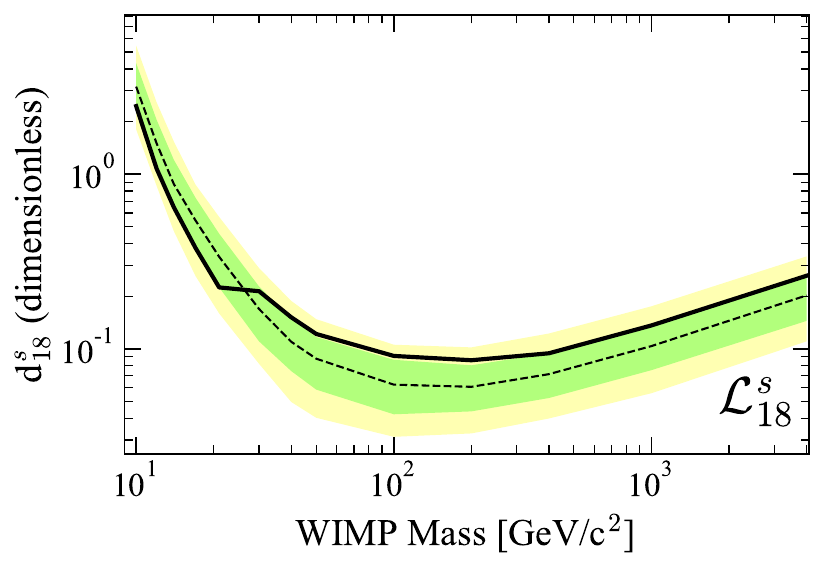}    
    \caption{
    The 90\% confidence limit (black lines) on the dimensionless isoscalar interaction couplings $d_j$ for each of the five interactions.
    The black dotted lines show the medians of the sensitivity projection, and the green and yellow bands correspond to the 1$\sigma$ and 2$\sigma$ sensitivity bands, respectively.
    Also shown are the results from PandaX-II experiment in blue where available.
    }
  \label{fig:limits-elastic-s}
\end{figure*}

\begin{figure*}
    \includegraphics[width=0.6\columnwidth]{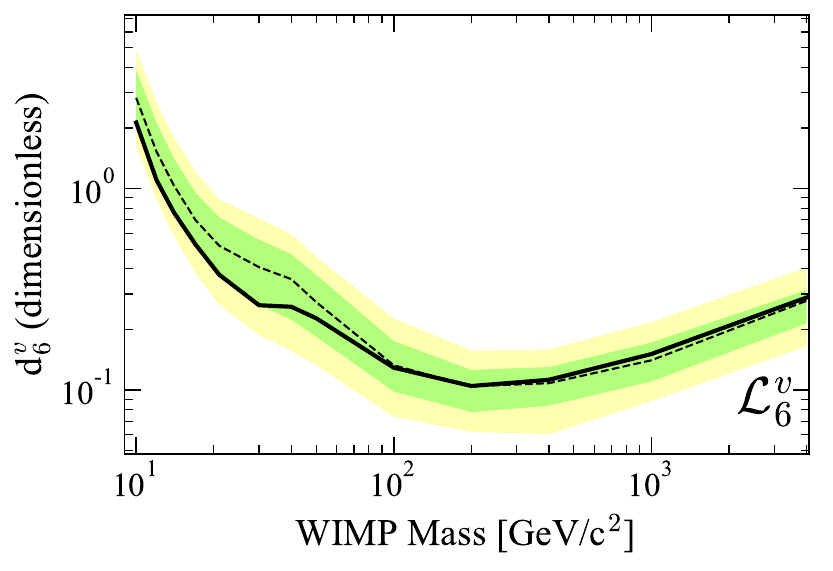}
    \includegraphics[width=0.6\columnwidth]{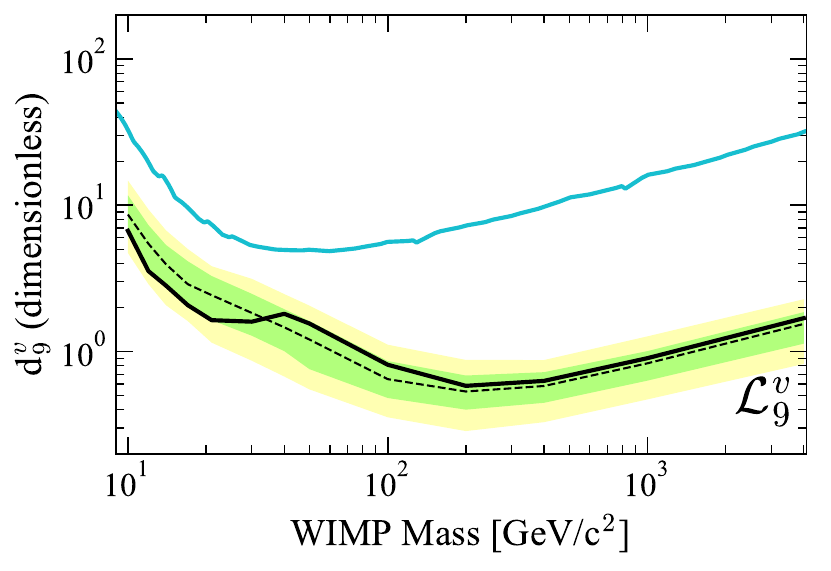}
    \includegraphics[width=0.6\columnwidth]{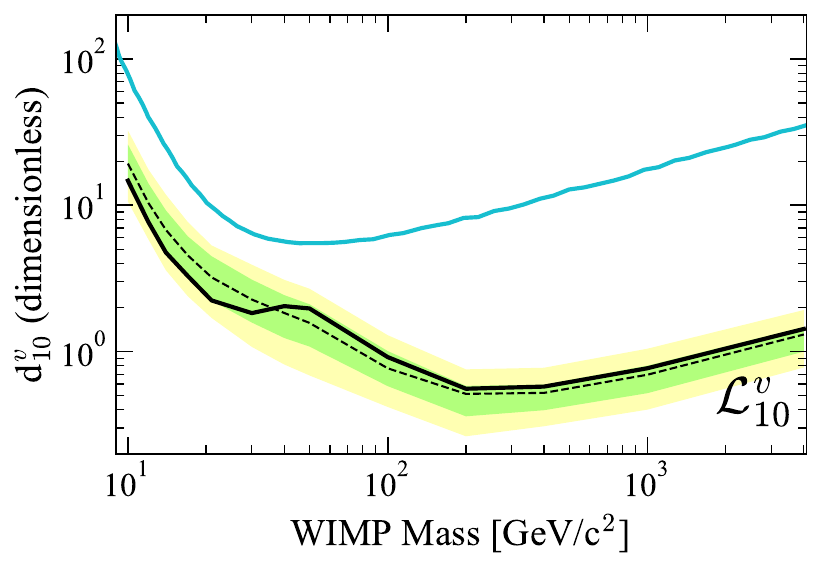}
    \includegraphics[width=0.6\columnwidth]{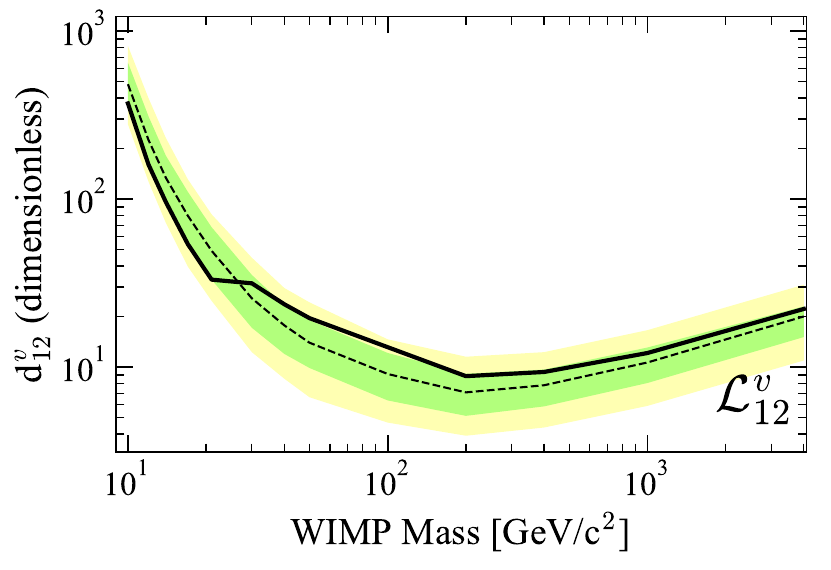}
    \includegraphics[width=0.6\columnwidth]{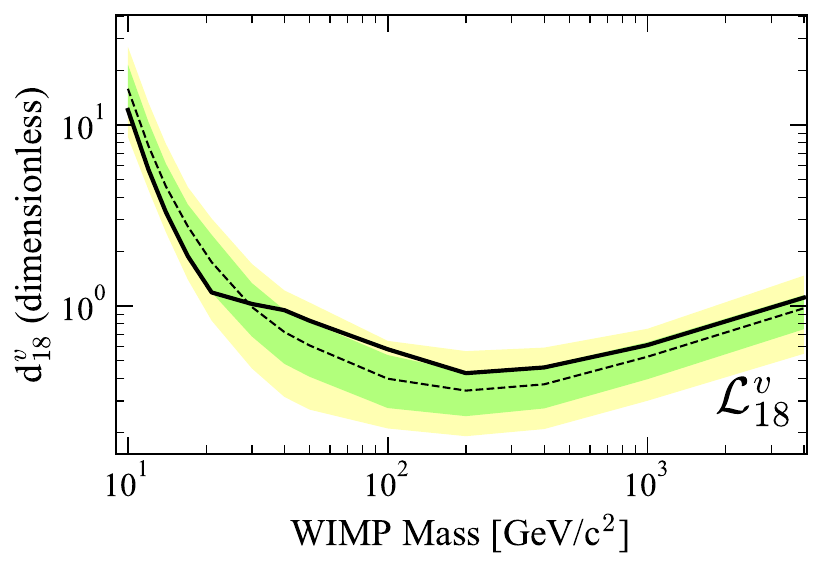}    
    \caption{
    The 90\% confidence limit (black lines) on the dimensionless isovector interaction couplings $d_j$ for each of the five interactions.
    The black dotted lines show the medians of the sensitivity projection, and the green and yellow bands correspond to the 1$\sigma$ and 2$\sigma$ sensitivity bands, respectively.
    Also shown are the results from PandaX-II experiment in blue where available.
    }
  \label{fig:limits-elastic-v}
\end{figure*}

No significant evidence of an excess is found in either the isoscalar or isovector bases.
Unbinned Kolmogorov-Smirnov tests comparing the reconstructed energy distributions of the data and the background-only model give p-values of 0.392.
This shows consistency with the background-only scenario for all Lagrangians and WIMP masses tested.
The interaction coupling parameters, $d_j$, for $\mathcal{L}_{6,9,10,12,18}$ are constrained, and shown in \autoref{fig:limits-elastic-s} and \autoref{fig:limits-elastic-v}.
Where available, previous limits on Lagrangians from PandaX-II~\cite{PandaX2:SD_EFT_2019} are shown for comparison. 
Each limit has an applied power constraint, employed to restrict the lower limit from dropping below 1$\sigma$ due to underfluctuations in the data. 
We restrict to an alternative hypothesis power of $\pi_{crit} = 0.16$.

The shape of each limit, and therefore the corresponding mass of maximum sensitivity, varies with the expected rate of events seen in the recoil spectra, shown in \autoref{fig:recoil_spectra}. 
Thus, Lagrangians with rates that increase at higher mass will have maximum sensitivity at higher mass. 
At masses above 30~GeV/c$^2$, some Lagrangians show limits weaker than the median expectation. 
This is due to the measured overfluctation of events in the NR band. These events can be seen as the data points that fall below the blue 2.5$\sigma$ contours of the ER band in~\autoref{fig:models_with_data}.
However, all of these events are consistent with ER leakage as described in Refs.~\cite{LZ:SR1WS_2022,LZ:SR1_NREFT_2023}.

The lower bounds of each Lagrangians are near, or below in the case of $\mathcal{L}_{6}$, the nominal weak scale where $d_j \approx 1$. 
These results give us information on the absolute size of the WIMP magnetic and electric dipole moment and their coupling to nucleons since we have normalized to the scale of the nucleon ($m_M = m_N$). 

\section{\label{sec:conclusion}Conclusion}
This letter presents the results of a search for covariant vector, electric dipole moment, and magnetic dipole moment interactions between a WIMP and a nucleon.
Ten different interaction nuclear recoil spectra were generated using relativistically matched NREFT operators.
Using a frequentist statistical analysis between data and model, no excess is observed for any model. 
Limits on the interaction coupling strength, $d_j$, were placed, for masses between 9~GeV/c$^2$ and 4000~GeV/c$^2$ for isoscalar and isovector interactions. 
This work places the strongest constraints to date for every model tested. 
These results help elucidate possible physics that may explain the behavior of the WIMP and its interactions with SM particles.

\begin{acknowledgments}
The research supporting this work took place in part at the Sanford Underground Research Facility (SURF) in Lead, South Dakota. 
Funding for this work is supported by the U.S. Department of Energy, Office of Science, Office of High Energy Physics under Contract Numbers DE-AC02-05CH11231, DE-SC0020216, DE-SC0012704, DE-SC0010010, DE-AC02-07CH11359, DE-SC0012161, DE-SC0015910, DE-SC0014223, DE-SC0010813, DE-SC0009999, DE-NA0003180, DE-SC0011702, DE-SC0010072, DE-SC0015708, DE-SC0006605, DE-SC0008475, DE-SC0019193, DE-FG02-10ER46709, UW PRJ82AJ, DE-SC0013542, DE-AC02-76SF00515, DE-SC0018982, DE-SC0019066, DE-SC0015535, DE-SC0019319, DE-SC0024225, DE-SC0024114, DE-AC52-07NA27344, \& DOE-SC0012447.
This research was also supported by U.S. National Science Foundation (NSF); the UKRI’s Science \& Technology Facilities Council under award numbers ST/M003744/1, ST/M003655/1, ST/M003639/1, ST/M003604/1, ST/M003779/1, ST/M003469/1, ST/M003981/1, ST/N000250/1, ST/N000269/1, ST/N000242/1, ST/N000331/1, ST/N000447/1, ST/N000277/1, ST/N000285/1, ST/S000801/1, ST/S000828/1, ST/S000739/1, ST/S000879/1, ST/S000933/1, ST/S000844/1, ST/S000747/1, ST/S000666/1, ST/R003181/1, ST/W000547/1, ST/W000636/1, ST/W000490/1; Portuguese Foundation for Science and Technology (FCT) under award numbers PTDC/FIS-PAR/2831/2020; the Institute for Basic Science, Korea (budget number IBS-R016-D1). 
We acknowledge additional support from the STFC Boulby Underground Laboratory in the U.K., the GridPP~\cite{faulkner2005gridpp,britton2009gridpp} and IRIS Collaborations, in particular at Imperial College London and additional support by the University College London (UCL) Cosmoparticle Initiative, and by the ARC Centre of Excellence for Dark Matter Particle Physics, and the University of Zurich.
We acknowledge additional support from the Center for the Fundamental Physics of the Universe, Brown University. 
K.T. Lesko acknowledges the support of Brasenose College and Oxford University. 
The LZ Collaboration acknowledges key contributions of Dr. Sidney Cahn, Yale University, in the production of calibration sources. 
This research used resources of the National Energy Research Scientific Computing Center, a DOE Office of Science User Facility supported by the Office of Science of the U.S. Department of Energy under Contract No. DE-AC02-05CH11231. We gratefully acknowledge support from GitLab through its GitLab for Education Program. 
The University of Edinburgh is a charitable body, registered in Scotland, with the registration number SC005336. 
The assistance of SURF and its personnel in providing physical access and general logistical and technical support is acknowledged. We acknowledge the South Dakota Governor's office, the South Dakota Community Foundation, the South Dakota State University Foundation, and the University of South Dakota Foundation for use of xenon.
We also acknowledge the University of Alabama for providing xenon.
For the purpose of open access, the authors have applied a Creative Commons Attribution (CC BY) licence to any Author Accepted Manuscript version arising from this submission.
\end{acknowledgments}

\bibliography{references}

\begin{thebibliography}{26}%
\makeatletter
\providecommand \@ifxundefined [1]{%
 \@ifx{#1\undefined}
}%
\providecommand \@ifnum [1]{%
 \ifnum #1\expandafter \@firstoftwo
 \else \expandafter \@secondoftwo
 \fi
}%
\providecommand \@ifx [1]{%
 \ifx #1\expandafter \@firstoftwo
 \else \expandafter \@secondoftwo
 \fi
}%
\providecommand \natexlab [1]{#1}%
\providecommand \enquote  [1]{``#1''}%
\providecommand \bibnamefont  [1]{#1}%
\providecommand \bibfnamefont [1]{#1}%
\providecommand \citenamefont [1]{#1}%
\providecommand \href@noop [0]{\@secondoftwo}%
\providecommand \href [0]{\begingroup \@sanitize@url \@href}%
\providecommand \@href[1]{\@@startlink{#1}\@@href}%
\providecommand \@@href[1]{\endgroup#1\@@endlink}%
\providecommand \@sanitize@url [0]{\catcode `\\12\catcode `\$12\catcode
  `\&12\catcode `\#12\catcode `\^12\catcode `\_12\catcode `\%12\relax}%
\providecommand \@@startlink[1]{}%
\providecommand \@@endlink[0]{}%
\providecommand \url  [0]{\begingroup\@sanitize@url \@url }%
\providecommand \@url [1]{\endgroup\@href {#1}{\urlprefix }}%
\providecommand \urlprefix  [0]{URL }%
\providecommand \Eprint [0]{\href }%
\providecommand \doibase [0]{http://dx.doi.org/}%
\providecommand \selectlanguage [0]{\@gobble}%
\providecommand \bibinfo  [0]{\@secondoftwo}%
\providecommand \bibfield  [0]{\@secondoftwo}%
\providecommand \translation [1]{[#1]}%
\providecommand \BibitemOpen [0]{}%
\providecommand \bibitemStop [0]{}%
\providecommand \bibitemNoStop [0]{.\EOS\space}%
\providecommand \EOS [0]{\spacefactor3000\relax}%
\providecommand \BibitemShut  [1]{\csname bibitem#1\endcsname}%
\let\auto@bib@innerbib\@empty
%</preamble>
\bibitem [{\citenamefont {Akerib}\ \emph {et~al.}(2020)\citenamefont {Akerib}
  \emph {et~al.}}]{LZ:Experiment_2020}%
  \BibitemOpen
  \bibfield  {author} {\bibinfo {author} {\bibfnamefont {D.~S.}\ \bibnamefont
  {Akerib}} \emph {et~al.} (\bibinfo {collaboration} {LZ Collaboration}),\
  }\href {\doibase 10.1016/j.nima.2019.163047} {\bibfield  {journal} {\bibinfo
  {journal} {Nucl. Instrum. Meth. A}\ }\textbf {\bibinfo {volume} {953}},\
  \bibinfo {pages} {163047} (\bibinfo {year} {2020})},\ \Eprint
  {http://arxiv.org/abs/1910.09124} {arXiv:1910.09124 [physics.ins-det]}
  \BibitemShut {NoStop}%
\bibitem [{\citenamefont {Aprile}\ \emph {et~al.}(2023)\citenamefont {Aprile}
  \emph {et~al.}}]{XenonNT:WS_2023}%
  \BibitemOpen
  \bibfield  {author} {\bibinfo {author} {\bibfnamefont {E.}~\bibnamefont
  {Aprile}} \emph {et~al.} (\bibinfo {collaboration} {XENON Collaboration}),\
  }\href {\doibase 10.1103/PhysRevLett.131.041003} {\bibfield  {journal}
  {\bibinfo  {journal} {Phys. Rev. Lett.}\ }\textbf {\bibinfo {volume} {131}},\
  \bibinfo {pages} {041003} (\bibinfo {year} {2023})},\ \Eprint
  {http://arxiv.org/abs/2303.14729} {arXiv:2303.14729 [hep-ex]} \BibitemShut
  {NoStop}%
\bibitem [{\citenamefont {Meng}\ \emph {et~al.}(2021)\citenamefont {Meng} \emph
  {et~al.}}]{PandaX4T:SI2023}%
  \BibitemOpen
  \bibfield  {author} {\bibinfo {author} {\bibfnamefont {Y.}~\bibnamefont
  {Meng}} \emph {et~al.} (\bibinfo {collaboration} {PandaX-4T Collaboration}),\
  }\href {\doibase 10.1103/PhysRevLett.127.261802} {\bibfield  {journal}
  {\bibinfo  {journal} {Phys. Rev. Lett.}\ }\textbf {\bibinfo {volume} {127}},\
  \bibinfo {pages} {261802} (\bibinfo {year} {2021})},\ \Eprint
  {http://arxiv.org/abs/2107.13438} {arXiv:2107.13438 [hep-ex]} \BibitemShut
  {NoStop}%
\bibitem [{\citenamefont {Aalbers}\ \emph
  {et~al.}(2023{\natexlab{a}})\citenamefont {Aalbers} \emph
  {et~al.}}]{LZ:SR1WS_2022}%
  \BibitemOpen
  \bibfield  {author} {\bibinfo {author} {\bibfnamefont {J.}~\bibnamefont
  {Aalbers}} \emph {et~al.} (\bibinfo {collaboration} {LZ Collaboration}),\
  }\href {\doibase 10.1103/PhysRevLett.131.041002} {\bibfield  {journal}
  {\bibinfo  {journal} {Phys. Rev. Lett.}\ }\textbf {\bibinfo {volume} {131}},\
  \bibinfo {pages} {041002} (\bibinfo {year} {2023}{\natexlab{a}})},\ \Eprint
  {http://arxiv.org/abs/2207.03764} {arXiv:2207.03764 [hep-ex]} \BibitemShut
  {NoStop}%
\bibitem [{\citenamefont {Fan}\ \emph {et~al.}(2010)\citenamefont {Fan},
  \citenamefont {Reece},\ and\ \citenamefont {Wang}}]{Fan_2010}%
  \BibitemOpen
  \bibfield  {author} {\bibinfo {author} {\bibfnamefont {J.}~\bibnamefont
  {Fan}}, \bibinfo {author} {\bibfnamefont {M.}~\bibnamefont {Reece}}, \ and\
  \bibinfo {author} {\bibfnamefont {L.-T.}\ \bibnamefont {Wang}},\ }\href
  {\doibase 10.1088/1475-7516/2010/11/042} {\bibfield  {journal} {\bibinfo
  {journal} {J. Cosmol. Astropart. Phys.}\ }\textbf {\bibinfo {volume}
  {2010}},\ \bibinfo {pages} {042} (\bibinfo {year} {2010})}\BibitemShut
  {NoStop}%
\bibitem [{\citenamefont {Fitzpatrick}\ \emph {et~al.}(2013)\citenamefont
  {Fitzpatrick}, \citenamefont {Haxton}, \citenamefont {Katz}, \citenamefont
  {Lubbers},\ and\ \citenamefont {Xu}}]{Fitzpatrick:EFT}%
  \BibitemOpen
  \bibfield  {author} {\bibinfo {author} {\bibfnamefont {A.~L.}\ \bibnamefont
  {Fitzpatrick}}, \bibinfo {author} {\bibfnamefont {W.}~\bibnamefont {Haxton}},
  \bibinfo {author} {\bibfnamefont {E.}~\bibnamefont {Katz}}, \bibinfo {author}
  {\bibfnamefont {N.}~\bibnamefont {Lubbers}}, \ and\ \bibinfo {author}
  {\bibfnamefont {Y.}~\bibnamefont {Xu}},\ }\href {\doibase
  10.1088/1475-7516/2013/02/004} {\bibfield  {journal} {\bibinfo  {journal}
  {JCAP}\ }\textbf {\bibinfo {volume} {1302}},\ \bibinfo {pages} {004}
  (\bibinfo {year} {2013})},\ \Eprint {http://arxiv.org/abs/1203.3542}
  {arXiv:1203.3542 [hep-ph]} \BibitemShut {NoStop}%
%%CITATION = ARXIV:1203.3542;%%
\bibitem [{\citenamefont {Akerib}\ \emph {et~al.}(2021)\citenamefont {Akerib}
  \emph {et~al.}}]{LUX:EFTR4_2021}%
  \BibitemOpen
  \bibfield  {author} {\bibinfo {author} {\bibfnamefont {D.}~\bibnamefont
  {Akerib}} \emph {et~al.} (\bibinfo {collaboration} {LUX Collaboration}),\
  }\href {\doibase 10.1103/physrevd.104.062005} {\bibfield  {journal} {\bibinfo
   {journal} {Phys. Rev. D}\ }\textbf {\bibinfo {volume} {104}} (\bibinfo
  {year} {2021}),\ 10.1103/physrevd.104.062005}\BibitemShut {NoStop}%
\bibitem [{\citenamefont {Aprile}\ \emph {et~al.}(2019)\citenamefont {Aprile}
  \emph {et~al.}}]{Xenon1t:2vec_2019}%
  \BibitemOpen
  \bibfield  {author} {\bibinfo {author} {\bibfnamefont {E.}~\bibnamefont
  {Aprile}} \emph {et~al.} (\bibinfo {collaboration} {XENON Collaboration}),\
  }\href {\doibase 10.1038/s41586-019-1124-4} {\bibfield  {journal} {\bibinfo
  {journal} {Nature}\ }\textbf {\bibinfo {volume} {568}},\ \bibinfo {pages}
  {532} (\bibinfo {year} {2019})}\BibitemShut {NoStop}%
\bibitem [{\citenamefont {Xia}\ \emph {et~al.}(2019)\citenamefont {Xia},
  \citenamefont {Haxton} \emph {et~al.}}]{PandaX2:SD_EFT_2019}%
  \BibitemOpen
  \bibfield  {author} {\bibinfo {author} {\bibfnamefont {J.}~\bibnamefont
  {Xia}}, \bibinfo {author} {\bibfnamefont {W.~C.}\ \bibnamefont {Haxton}},
  \emph {et~al.} (\bibinfo {collaboration} {PandaX Collaboration}),\ }\href
  {\doibase 10.1016/j.physletb.2019.02.043} {\bibfield  {journal} {\bibinfo
  {journal} {Phys. Lett. B}\ }\textbf {\bibinfo {volume} {792}},\ \bibinfo
  {pages} {193} (\bibinfo {year} {2019})}\BibitemShut {NoStop}%
\bibitem [{\citenamefont {Aalbers}\ \emph
  {et~al.}(2023{\natexlab{b}})\citenamefont {Aalbers} \emph
  {et~al.}}]{LZ:SR1_NREFT_2023}%
  \BibitemOpen
  \bibfield  {author} {\bibinfo {author} {\bibfnamefont {J.}~\bibnamefont
  {Aalbers}} \emph {et~al.} (\bibinfo {collaboration} {LZ Collaboration}),\
  }\href {\doibase 10.1103/PhysRevLett.131.041002} {\bibfield  {journal}
  {\bibinfo  {journal} {Phys. Rev. Lett.}\ }\textbf {\bibinfo {volume} {131}},\
  \bibinfo {pages} {041002} (\bibinfo {year} {2023}{\natexlab{b}})},\ \Eprint
  {http://arxiv.org/abs/2207.03764} {arXiv:2207.03764 [hep-ex]} \BibitemShut
  {NoStop}%
\bibitem [{\citenamefont {Anand}\ \emph {et~al.}(2014)\citenamefont {Anand},
  \citenamefont {Fitzpatrick},\ and\ \citenamefont
  {Haxton}}]{Anand:MathematicaEFT}%
  \BibitemOpen
  \bibfield  {author} {\bibinfo {author} {\bibfnamefont {N.}~\bibnamefont
  {Anand}}, \bibinfo {author} {\bibfnamefont {A.~L.}\ \bibnamefont
  {Fitzpatrick}}, \ and\ \bibinfo {author} {\bibfnamefont {W.~C.}\ \bibnamefont
  {Haxton}},\ }\href {\doibase 10.1103/PhysRevC.89.065501} {\bibfield
  {journal} {\bibinfo  {journal} {Phys. Rev. C}\ }\textbf {\bibinfo {volume}
  {89}},\ \bibinfo {pages} {065501} (\bibinfo {year} {2014})}\BibitemShut
  {NoStop}%
\bibitem [{\citenamefont {Agrawal}\ \emph {et~al.}(2021)\citenamefont {Agrawal}
  \emph {et~al.}}]{Agrawal:2021dbo}%
  \BibitemOpen
  \bibfield  {author} {\bibinfo {author} {\bibfnamefont {P.}~\bibnamefont
  {Agrawal}} \emph {et~al.},\ }\href {\doibase 10.1140/epjc/s10052-021-09703-7}
  {\bibfield  {journal} {\bibinfo  {journal} {Eur. Phys. J. C}\ }\textbf
  {\bibinfo {volume} {81}},\ \bibinfo {pages} {1015} (\bibinfo {year}
  {2021})},\ \Eprint {http://arxiv.org/abs/2102.12143} {arXiv:2102.12143
  [hep-ph]} \BibitemShut {NoStop}%
\bibitem [{\citenamefont {Mount}\ \emph {et~al.}(2017)\citenamefont {Mount}
  \emph {et~al.}}]{LZ:TDR_2017}%
  \BibitemOpen
  \bibfield  {author} {\bibinfo {author} {\bibfnamefont {B.~J.}\ \bibnamefont
  {Mount}} \emph {et~al.} (\bibinfo {collaboration} {LZ Collaboration}),\
  }\href@noop {} {\  (\bibinfo {year} {2017})},\ \Eprint
  {http://arxiv.org/abs/1703.09144} {arXiv:1703.09144 [physics.ins-det]}
  \BibitemShut {NoStop}%
\bibitem [{\citenamefont {Szydagis}\ \emph
  {et~al.}(2022{\natexlab{a}})\citenamefont {Szydagis} \emph
  {et~al.}}]{NEST:paper_2022}%
  \BibitemOpen
  \bibfield  {author} {\bibinfo {author} {\bibfnamefont {M.}~\bibnamefont
  {Szydagis}} \emph {et~al.} (\bibinfo {collaboration} {NEST Collaboration}),\
  }\href {\doibase 10.5281/zenodo.6534007} {\enquote {\bibinfo {title} {Noble
  element simulation technique},}\ } (\bibinfo {year}
  {2022}{\natexlab{a}})\BibitemShut {NoStop}%
\bibitem [{\citenamefont {Szydagis}\ \emph
  {et~al.}(2022{\natexlab{b}})\citenamefont {Szydagis} \emph
  {et~al.}}]{NEST:paper_2023}%
  \BibitemOpen
  \bibfield  {author} {\bibinfo {author} {\bibfnamefont {M.}~\bibnamefont
  {Szydagis}} \emph {et~al.},\ }\href@noop {} {\  (\bibinfo {year}
  {2022}{\natexlab{b}})},\ \Eprint {http://arxiv.org/abs/2211.10726}
  {arXiv:2211.10726 [hep-ex]} \BibitemShut {NoStop}%
\bibitem [{\citenamefont {Jeong}\ \emph {et~al.}(2022)\citenamefont {Jeong},
  \citenamefont {Kang}, \citenamefont {Scopel},\ and\ \citenamefont
  {Tomar}}]{Jeong_2022}%
  \BibitemOpen
  \bibfield  {author} {\bibinfo {author} {\bibfnamefont {I.}~\bibnamefont
  {Jeong}}, \bibinfo {author} {\bibfnamefont {S.}~\bibnamefont {Kang}},
  \bibinfo {author} {\bibfnamefont {S.}~\bibnamefont {Scopel}}, \ and\ \bibinfo
  {author} {\bibfnamefont {G.}~\bibnamefont {Tomar}},\ }\href {\doibase
  10.1016/j.cpc.2022.108342} {\bibfield  {journal} {\bibinfo  {journal}
  {Comput. Phys. Commun.}\ }\textbf {\bibinfo {volume} {276}},\ \bibinfo
  {pages} {108342} (\bibinfo {year} {2022})}\BibitemShut {NoStop}%
\bibitem [{\citenamefont {Menéndez}\ \emph {et~al.}(2009)\citenamefont
  {Menéndez}, \citenamefont {Poves}, \citenamefont {Caurier},\ and\
  \citenamefont {Nowacki}}]{MENENDEZ2009139}%
  \BibitemOpen
  \bibfield  {author} {\bibinfo {author} {\bibfnamefont {J.}~\bibnamefont
  {Menéndez}}, \bibinfo {author} {\bibfnamefont {A.}~\bibnamefont {Poves}},
  \bibinfo {author} {\bibfnamefont {E.}~\bibnamefont {Caurier}}, \ and\
  \bibinfo {author} {\bibfnamefont {F.}~\bibnamefont {Nowacki}},\ }\href
  {\doibase https://doi.org/10.1016/j.nuclphysa.2008.12.005} {\bibfield
  {journal} {\bibinfo  {journal} {Nucl. Phys. A}\ }\textbf {\bibinfo {volume}
  {818}},\ \bibinfo {pages} {139} (\bibinfo {year} {2009})}\BibitemShut
  {NoStop}%
\bibitem [{\citenamefont {Fitzpatrick}\ \emph {et~al.}()\citenamefont
  {Fitzpatrick}, \citenamefont {Haxton}, \citenamefont {Johnson},\ and\
  \citenamefont {McElvain}}]{haxton_unpublished}%
  \BibitemOpen
  \bibfield  {author} {\bibinfo {author} {\bibfnamefont {A.~L.}\ \bibnamefont
  {Fitzpatrick}}, \bibinfo {author} {\bibfnamefont {W.~C.}\ \bibnamefont
  {Haxton}}, \bibinfo {author} {\bibfnamefont {C.~W.}\ \bibnamefont {Johnson}},
  \ and\ \bibinfo {author} {\bibfnamefont {K.~S.}\ \bibnamefont {McElvain}},\
  }\href@noop {} {\bibinfo  {journal} {in preparation for {Annu. Rev. Nucl.
  Part. Sci}}\ }\BibitemShut {NoStop}%
\bibitem [{\citenamefont {Baxter}\ \emph {et~al.}(2021)\citenamefont {Baxter}
  \emph {et~al.}}]{DM_parameters:BAXTER2021_Conventions}%
  \BibitemOpen
\bibfield  {journal} {  }\bibfield  {author} {\bibinfo {author} {\bibfnamefont
  {D.}~\bibnamefont {Baxter}} \emph {et~al.},\ }\href {\doibase
  10.1140/epjc/s10052-021-09655-y} {\bibfield  {journal} {\bibinfo  {journal}
  {Eur. Phys. J. C}\ }\textbf {\bibinfo {volume} {81}} (\bibinfo {year}
  {2021}),\ 10.1140/epjc/s10052-021-09655-y}\BibitemShut {NoStop}%
\bibitem [{\citenamefont {Schoenrich}\ \emph {et~al.}(2010)\citenamefont
  {Schoenrich}, \citenamefont {Binney},\ and\ \citenamefont
  {Dehnen}}]{Schoenrich:Local_kinematics}%
  \BibitemOpen
  \bibfield  {author} {\bibinfo {author} {\bibfnamefont {R.}~\bibnamefont
  {Schoenrich}}, \bibinfo {author} {\bibfnamefont {J.}~\bibnamefont {Binney}},
  \ and\ \bibinfo {author} {\bibfnamefont {W.}~\bibnamefont {Dehnen}},\ }\href
  {\doibase 10.1111/j.1365-2966.2010.16253.x} {\bibfield  {journal} {\bibinfo
  {journal} {Mon. Not. Roy. Astron. Soc.}\ }\textbf {\bibinfo {volume} {403}},\
  \bibinfo {pages} {1829} (\bibinfo {year} {2010})},\ \Eprint
  {http://arxiv.org/abs/0912.3693} {arXiv:0912.3693 [astro-ph.GA]} \BibitemShut
  {NoStop}%
\bibitem [{\citenamefont {Bland-Hawthorn}\ and\ \citenamefont
  {Gerhard}(2016)}]{DM_parameters:galaxy_context_rest_velocity_1}%
  \BibitemOpen
  \bibfield  {author} {\bibinfo {author} {\bibfnamefont {J.}~\bibnamefont
  {Bland-Hawthorn}}\ and\ \bibinfo {author} {\bibfnamefont {O.}~\bibnamefont
  {Gerhard}},\ }\href {\doibase 10.1146/annurev-astro-081915-023441} {\bibfield
   {journal} {\bibinfo  {journal} {Annu. Rev. Astron. Astrophys.}\ }\textbf
  {\bibinfo {volume} {54}},\ \bibinfo {pages} {529} (\bibinfo {year} {2016})},\
  \Eprint
  {http://arxiv.org/abs/https://doi.org/10.1146/annurev-astro-081915-023441}
  {https://doi.org/10.1146/annurev-astro-081915-023441} \BibitemShut {NoStop}%
\bibitem [{\citenamefont {{Abuter, R.}}\ \emph {et~al.}(2021)\citenamefont
  {{Abuter, R.}} \emph
  {et~al.}}]{DM_parameters:galaxy_context_rest_velocity_2}%
  \BibitemOpen
  \bibfield  {author} {\bibinfo {author} {\bibnamefont {{Abuter, R.}}} \emph
  {et~al.} (\bibinfo {collaboration} {{GRAVITY Collaboration}}),\ }\href
  {\doibase 10.1051/0004-6361/202040208} {\bibfield  {journal} {\bibinfo
  {journal} {A\&A}\ }\textbf {\bibinfo {volume} {647}},\ \bibinfo {pages} {A59}
  (\bibinfo {year} {2021})}\BibitemShut {NoStop}%
\bibitem [{\citenamefont {Smith}\ \emph {et~al.}(2007)\citenamefont {Smith}
  \emph {et~al.}}]{DM_parameters:RAVE_survey_escape_velocity}%
  \BibitemOpen
  \bibfield  {author} {\bibinfo {author} {\bibfnamefont {M.~C.}\ \bibnamefont
  {Smith}} \emph {et~al.},\ }\href {\doibase 10.1111/j.1365-2966.2007.11964.x}
  {\bibfield  {journal} {\bibinfo  {journal} {Mon. Not. Roy. Astron. Soc.}\
  }\textbf {\bibinfo {volume} {379}},\ \bibinfo {pages} {755} (\bibinfo {year}
  {2007})},\ \Eprint {http://arxiv.org/abs/astro-ph/0611671}
  {arXiv:astro-ph/0611671} \BibitemShut {NoStop}%
\bibitem [{\citenamefont {Lewin}\ and\ \citenamefont
  {Smith}(1996)}]{DM_parameters:LEWIN199687_DM_density}%
  \BibitemOpen
  \bibfield  {author} {\bibinfo {author} {\bibfnamefont {J.}~\bibnamefont
  {Lewin}}\ and\ \bibinfo {author} {\bibfnamefont {P.}~\bibnamefont {Smith}},\
  }\href {\doibase https://doi.org/10.1016/S0927-6505(96)00047-3} {\bibfield
  {journal} {\bibinfo  {journal} {Astropart. Phys.}\ }\textbf {\bibinfo
  {volume} {6}},\ \bibinfo {pages} {87} (\bibinfo {year} {1996})}\BibitemShut
  {NoStop}%
\bibitem [{\citenamefont {Faulkner}\ \emph {et~al.}(2005)\citenamefont
  {Faulkner} \emph {et~al.}}]{faulkner2005gridpp}%
  \BibitemOpen
  \bibfield  {author} {\bibinfo {author} {\bibfnamefont {P.}~\bibnamefont
  {Faulkner}} \emph {et~al.},\ }\href {\doibase
  https://doi.org/10.1088/0954-3899/32/1/N01} {\bibfield  {journal} {\bibinfo
  {journal} {J. Phys. G}\ }\textbf {\bibinfo {volume} {32}},\ \bibinfo {pages}
  {N1} (\bibinfo {year} {2005})}\BibitemShut {NoStop}%
\bibitem [{\citenamefont {Britton}\ \emph {et~al.}(2009)\citenamefont {Britton}
  \emph {et~al.}}]{britton2009gridpp}%
  \BibitemOpen
  \bibfield  {author} {\bibinfo {author} {\bibfnamefont {D.}~\bibnamefont
  {Britton}} \emph {et~al.},\ }\href {\doibase
  https://doi.org/10.1098/rsta.2009.0036} {\bibfield  {journal} {\bibinfo
  {journal} {Philos. Trans. R. Soc. A}\ }\textbf {\bibinfo {volume} {367}},\
  \bibinfo {pages} {2447} (\bibinfo {year} {2009})}\BibitemShut {NoStop}%
\end{thebibliography}%

\appendix
%\newpage
\section{Data Release\label{ap:2}}
Data from selected plots in this paper can be accessed at: \href{https://tinyurl.com/LZDataReleaseRun1HENR}{https://tinyurl.com/LZDataReleaseRun1HENR}.
The data available is:
\begin{itemize}
    \item[--] \hyperref[fig:models_with_data]{Figure~\ref*{fig:models_with_data}}: Points in S1-S2 space representing the data used in the WIMP search (black points).
    \item[--] \hyperref[fig:limits-elastic-s]{Figure~\ref*{fig:limits-elastic-s}}: Points representing the observed 90\% confidence level upper limits, together with the median, $\pm1\sigma$, and $\pm2\sigma$ expected sensitivities.
    \item[--] \hyperref[fig:limits-elastic-v]{Figure~\ref*{fig:limits-elastic-v}}: Points representing the observed 90\% confidence level upper limits, together with the median, $\pm1\sigma$, and $\pm2\sigma$ expected sensitivities.
    \item[--] Additionally, points representing the observed 90\% confidence level upper limits, together with the median, $\pm1\sigma$, and $\pm2\sigma$ expected sensitivities for $\mathcal{L}_{1-5}$, $\mathcal{L}_{7-8}$, $\mathcal{L}_{11}$, $\mathcal{L}_{13-17}$ and $\mathcal{L}_{19-20}$.
\end{itemize}

\end{document}